\documentclass[a4paper, 10pt, twocolumn]{article}

\usepackage{amsfonts,amssymb,amsmath,amsthm}
\usepackage{subfigure}
\usepackage{graphicx}
\usepackage[footnotesize]{caption}

\usepackage[top=1.5cm, left=1.5cm, right=1.5cm, bottom=1.5cm]{geometry}

\title{Some detection tests for low complexity data models \\ and unknown background distribution. }

\author{David Mary$^{1}$, S\'ebastien Bourguignon$^2$, \'Etienne Roquain$^{3}$,  \\ Sophia Sulis$^4$ and Marie Perrot-Dockes$^{3}$.\\ \footnotesize $^{1}$  Laboratoire Lagrange, Universit\'e C\^ote d'Azur, Observatoire de la C\^ote d'Azur.\\ \footnotesize $^{2}$ Laboratoire des Sciences du Num\'erique de Nantes,  \'Ecole Centrale de Nantes.\\ \footnotesize $^{3}$ Laboratoire de Probabilit\'es, Statistique et Mod\'elisation, Sorbonne Universit\'e.\\ \footnotesize $^{4}$ Laboratoire d'Astrophysique de Marseille, Aix Marseille Universit\'e.   }

\date{\empty} 

\renewenvironment{abstract}{\bf\small {\em\ Abstract---}}

\usepackage{amsmath}

\usepackage{hyperref,url}
\usepackage{graphicx}
\usepackage{tikz}
\usepackage{color}

\usepackage{bbold,amsmath,amstext,amsfonts,bbm,amssymb}

\begin{document}
	\maketitle
	\vspace*{-6mm}
		\begin{abstract}  We consider several detection situations  where, under the alternative hypothesis, the signal admits a low complexity model and, under both the null and the alternative hypotheses,  the distribution of the background noise is {unknown}. We present several detection strategies for such cases, whose design relies on exogenous or on endogenous data. These testing procedures have been inspired by and are applied to  two specific problems in Astrophysics, namely the detection of exoplanets from radial  velocity curves and of distant galaxies in hyperspectral datacubes.   
	\end{abstract}
	\vspace{-4mm}
	\section{Introduction}
	\label{sec:introduction}
	\vspace{-2mm}
	Several fields like Genomics or Astrophysics often require to perform a large number of statistical tests simultaneously. 
	During the last two decades, these applications have favored the emergence of theoretical works in multiple testing. Low-complexity- or sparsity-adaptive tests have been designed, including for instance the  Benjamini-Hochberg \cite{BH1995} procedure, which controls the  \textit{ false discovery rate} (FDR), or several variants  of Donoho and Jin's \textit{higher criticism} (HC)  \cite{Donoho_2004} and of the Berk-Jones (BJ) tests \cite{Moscovich_2016,Aldor_2013,Mary_2014,Kaplan_2014,Gontscharuk_2014}.
	In practice however, the distribution of the noise background
	may not be well known or even unknown. This may explain why such multiple testing procedures, which require the knowledge of the distribution under the null hypothesis, are not used 
	in practical applications where they would yet be useful. Users may prefer in practice
	considering more ad hoc procedures with the hope, sometimes illusory, that the test is performed at a controlled level. A recent example in Astrophysics is the claimed exoplanet detection around  $\alpha$ Centauri Bb, with a P-value ({\it i.e.}, the probability, {under ${\cal H}_0$}, of observing a test statistic at least as extreme as the one observed) evaluated at $0.02\% $ \cite{Dumusque_2012}. This detection was strongly discarded by further analyses \cite{Hatzes_2013, Rajpaul_2016}, stressing out
	the necessity of considering effects that are difficult to control, such as the noise background created by the star hosting the planet. 
	
	This papers presents some testing procedures with correctly controlled error risks, even when the distribution of the test statistics is not well known. We consider two cases: exoplanet detection from radial velocity data \cite{sulis:tel-01687077} and  the detection of distant galaxies in the integral field spectrograph  MUSE installed at the Very Large Telescope in Chile \cite{bacon2017, Mary_2020}.

	\section{Exoplanet detection from time series}
	\label{sec2}
	The presence of a planet orbiting a star induces a reflex motion of the star around the star-planet mass centre. This motion  induces a quasi-periodic modulation of the radial velocity of the star with respect to a distant observer and imprints its signature on the stellar spectrum by the Doppler effect \cite{Perryman_2018}. By measuring the Doppler shift in the lines of the stellar spectrum, one obtains the radial velocity of the star at a given time. Data thus consist of a time series containing such measurements.
	
	%
	One way to setup the detection problem considers the following {\it composite} {({\it i.e.,} with unknown parameters)} hypotheses: 
	\begin{equation*} 
	\left\{         
	\begin{aligned}
	\text{ ${\cal{H}}_0$ : } {X}(t_j) &= {\displaystyle{{{E}}}}(t_j) \\
	\text{ ${\cal{H}}_1$ : } X(t_j) &=  \sum_{q = 1}^{N_s} \alpha_q \sin(2\pi f_q t_j+\varphi_q)+ E(t_j) \\
	\end{aligned}
	\right.
	\label{hyp}
	\end{equation*}
	%
	where $X(t_j=j\Delta t),\; j=1,\hdots,N$, are the data points and $E(t_j)$ is a zero-mean second-order stationary Gaussian noise. The signal under the alternative hypothesis ${\cal H}_1$  has low complexity since it is modeled by a few sine waves, in some unknown number  and with unknown parameters.
	That is, the  planetary signature is sparse in the Fourier space. 
	Under both hypotheses, the noise covariance matrix $\Sigma$  (or equivalently the power spectrum density $S$) is unknown, 
	owing to the adversarial presence of {correlated} stellar noise. {As a consequence,	the distribution of the periodogram $P$ at each frequency $\nu$, which depends on $S$,  is also unknown. For instance, under $\mathcal{H}_0$, $P(\nu)$ is (asymptotically) a  $\chi^2$ random variable scaled by $S(\nu)$:   $P(\nu \; ;\; {\mathcal{H}_0}) \sim {S(\nu)}\chi^2_2/2$}.
	
	To circumvent this difficulty we propose an {\it exogenous} approach, {{\it i.e.,}} based on ancillary data  under $H_0$. {These data} are simulated time series corresponding to stellar noise, available {\it via} astrophysical simulations (see our recent work~\cite{Sulis20203DMS}). 
	The computation load of such simulations is however very demanding, so the number (say, $L$) of training time series is small ({up to a few tens, typically}).

	The considered framework is also relevant in other contexts. 
	For instance, in asteroseismology, one wants to detect periodic signals in photometric time series, 
	where the noise is mainly due to atmospheric perturbations. In modern photometers, auxiliary optical channels are often devoted to monitor ``empty''  regions of the sky and stable calibration stars \cite{Gupta_2001}. 
	Note  that radar systems also use training noise samples  for detection since ages. However,  an important difference is that adaptive test statistics  in radar typically rely on the estimation of the noise covariance matrix 
	and therefore require $L>N$ \cite{Trees_2002}. This is a very different regime from that considered here, where $L\ll N$.
	
	The method simply consists in  building an averaged periodogram {(say, $\overline{P}_L(\nu)$)}  from the $L$ training time series. {Then, the standardized periodogram is defined as $Z(\nu):=P(\nu)/\overline{P}_L(\nu)$}.   Let $Z_1,\dots,Z_N$ denote {a subset of} its components.  For evenly sampled data, the  distributions {of $Z$ } admit analytical expressions {under both hypotheses} 
	\cite{ieee_sophia}. Under {standard} assumptions \cite{Li_2014},
	one can show that $Z_1,\dots,Z_N$ are asymptotically independent with $Z_j\sim \mathcal{F}(2,2L)$ (a Fisher F-distribution with parameters $2$ and $2L$) under the null hypothesis and $Z_j\sim \mathcal{F}_{\lambda_j}(2,2L)$ (non central F-distribution),
	whose non-centrality parameters $\lambda_j$ depend on the planetary signature. 
	
	Under $H_0$, the distribution of the standardized periodogram  is independent of $S$. 
	This property, together with asymptotic independence,  allows one to build global hypothesis tests of level $\alpha$ from values of $Z_j$, {\it e.g.,} with the test of the maximum:  this test rejects the null  hypothesis if  $\max(Z_j,1\leq j\leq N) $ is larger than a threshold $c_\alpha$, calibrated so that \cite{ieee_sophia} 
	$$
	1- \Big( 1-\Big(\frac{L}{c_\alpha+L}\Big)^L\Big)^{N}=\alpha.
	$$
	Similar results can be obtained for several
		other tests like that of  Fisher \cite{Fisher_1929} or its variants \cite{Chiu_1989, Shimshoni_1971,Bolviken_1983a}. This also allows one
			to use approaches such as   HC et BJ in a context where the  exact distribution under ${\mathcal{H}_0}$ is not known.
	
\begin{figure}[h] 
		\begin{tabular}{cc}
			\includegraphics [height=4cm]{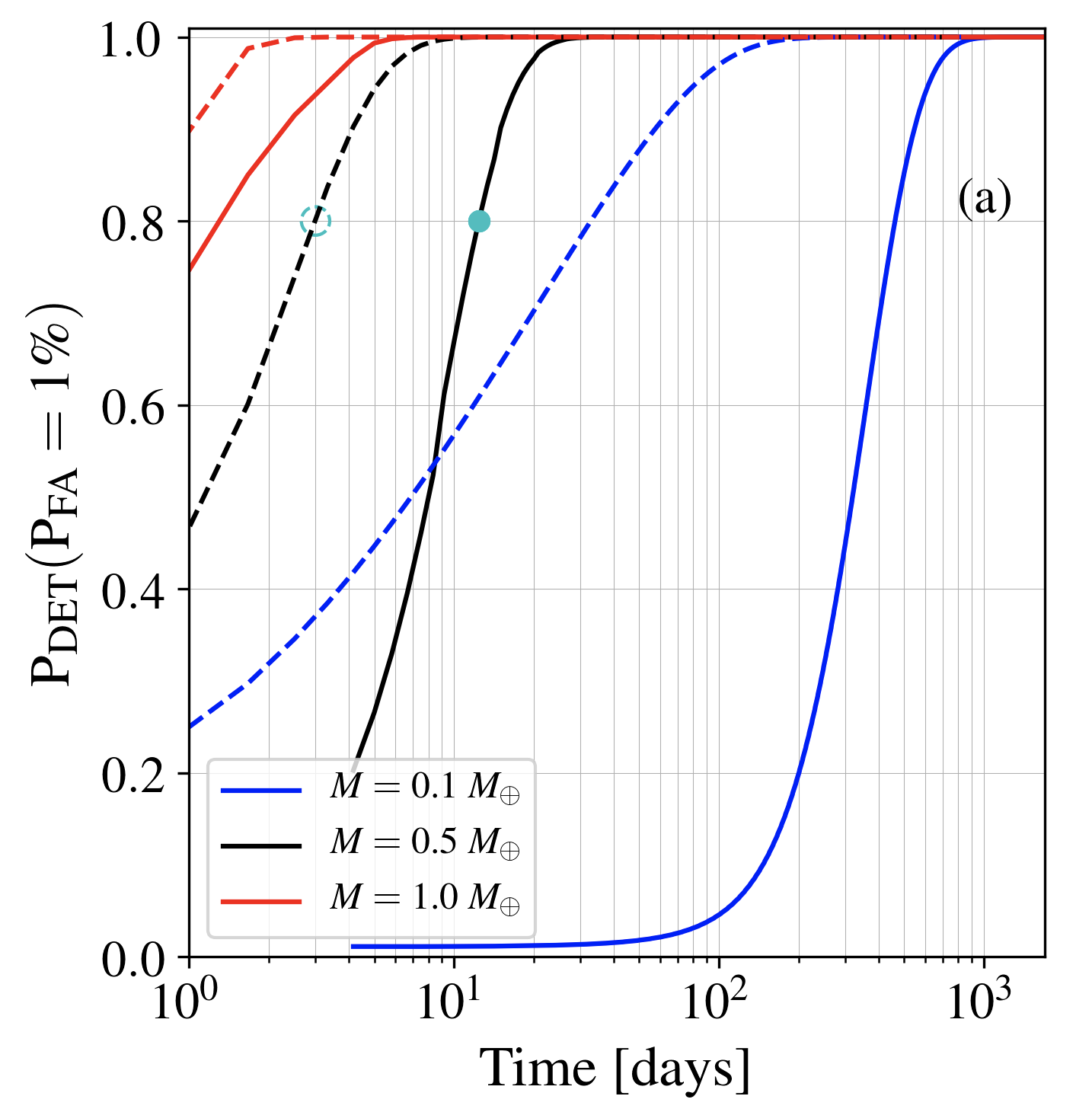} & \includegraphics[height=4.5cm,width=4.5cm]{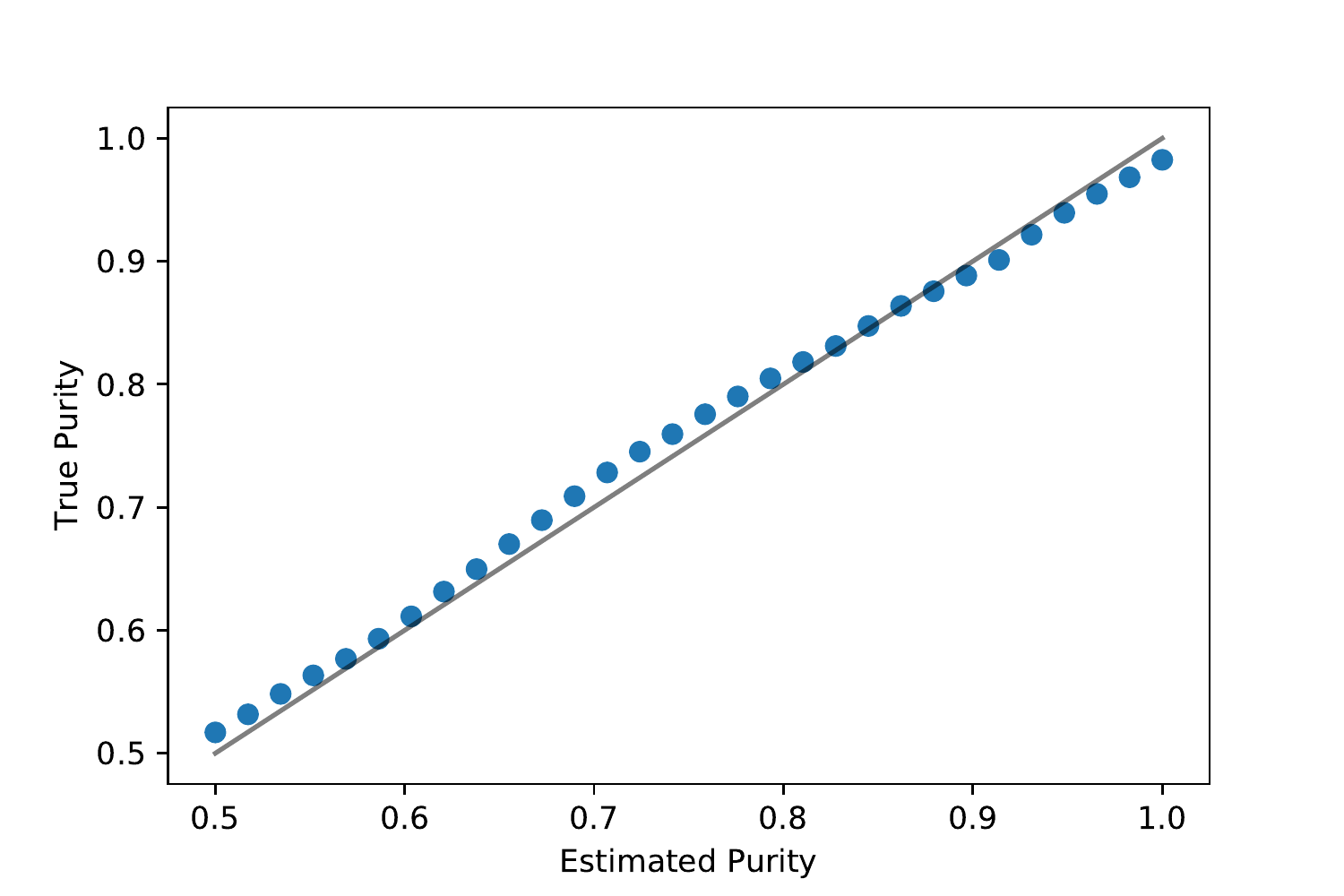}\end{tabular}
		\caption{{\bf Left panel: exoplanet detection} (Section~\ref{sec2}).: Power of the detection test for a fixed false alarm rate of $1\%$ as a function of the observing duration, for a planet orbiting circularly a solar-type star in 17.5 h. The considered exoplanet masses are 
			$0.1$ (blue), {0.5} (black) and {1} (red) Earth mass. The sampling step is $2$ hours and $L=20$. Solid lines correspond to correlated noise (realistic case)	and dashed lines to white noise. The noise standard deviation is $49$ cm.s$^{-1}$ in both cases.
			The planetary signal amplitude typically lies in the range $[1\; 10]$ cm.s$^{-1}$, depending on the planet mass. The detectability decreases for correlated noise because the frequency of the searched signal lies in a region where the stellar noise is stronger. In the considered configuration, the detection of a planet with half an Earth mass would be possible with probability  $80\%$ if observed during $12.4$ days. With white noise, this duration would be of only $3$ days (blue disk vs. circle). 			
			{\bf Right panel: galaxy detection} (Section~\ref{sec3}). True purity versus  purity estimated by the proposed method, for the emission line detection algorithm. The purity is defined as the fraction of true discoveries (true emission lines) over the total number of discoveries. The purity estimation method is evaluated on simulated MUSE data.}
		\label{fig1}
	\end{figure}

	In simulations, we have shown that the asymptotic approximation is accurate when the length of the time series is sufficiently large with respect to the	typical correlation time of the noise, and that the tests are conducted at the target false alarm rates, with the predicted probabilities of detection. Moreover, the analytical models allow to study the asymptotic powers of the tests, whose expressions are also accurate for moderate values of $N$. Such results allow for  detectability studies combining specific planetary parameters and instrumental configurations \cite{sulis:tel-01687077}.  An example is shown in the left panel of
	Fig.~\ref{fig1}, for the test defined by the maximum of the periodogram \cite{sulis:tel-01687077}. 

	\section{Detection of galaxies in MUSE}
	\label{sec3}
	The integral field spectrograph MUSE, installed on one
	of the 8-meter telescopes of the Very Large Telescope (Chile) acquires very rich hyperspectral images  (typically $300\times300$  pixels in 3\,600 optical channels). In such a  data ``cube'', we seek to detect very distant and faint galaxies, whose signatures essentially consist of a weak  increase of the light flux corresponding to an emission line, 
	convolved by the spatial point spread  function of the instrument.
	The shape of the lines is approximately known, but
	not their amplitudes, their positions in the data cube,  and nor their number---a few hundreds typically, against several hundreds of millions of voxels. In practice, the lines amplitudes can be very weak with respect to other astrophysical sources and to the background level, caused by the detector noise, residual sky emission, instrumental artifacts, etc.

	In this framework, we have considered a detection approach in three steps : the nuisance sources are first eliminated by linear subspace filtering techniques;  a specific matched filtering step is applied to the residual; the detection is performed in the filtered data residuals.   
	In order to focus on the sparse and localized nature of the target signatures, the considered test statistics are the local maxima of the resulting filtered data cube.  Because of the numerous and complex preprocessing steps applied to the data before the detection step,  the distribution of the test statistics under $H_0$ is unknown, with little hope to ever establish a reliable theoretical characterization.

	{In the considered problem }  there are several null hypotheses, each one linked to the position of a local maximum  \cite{cheng2017}. 
	If we denote by   $x,y,z$ the position of a local maximum ($x$ and $y$ are spatial coordinates, while $z$ is the wavelength channel), we consider the two hypotheses :
	\begin{equation*} 
	\left\{         
	\begin{aligned}
	\text{ $H_{0,x,y,z}$ : \text{there is no emission line at  position}  $(x,y,z)$},  \\
	\text{ $H_{1,x,y,z}$ : \text{ there is one such line at this position.} \;\;\;\;\;\;\;\;\;} \\
	\end{aligned}
	\right.
	\end{equation*}    
	The error criterion considered here is that of the FDR, that is, the expected value of the proportion of false discoveries over  the total number of positions found as possibly corresponding to a line \cite{BH1995, Mary_2020}.
	In contrast to the previous application, the approach considered here is {\it endogenous}, in the sense that it does not rely on auxiliary data. We have shown on simulated data that  the distribution of the local maxima under the joint null hypothesis ({\it i.e.}, there is no line in the whole data cube) can be estimated from the data itself, and that the whole procedure (including nuisance cancellation and matched filtering) controls the FDR (see Fig. \ref{fig1}, right panel). A detailed presentation of astrophysical results obtained with this algorithm can be found in \cite{Mary_2020}.

	\section{Conclusions and perspectives}
	In the signal processing community, there is a large literature on constant false alarm rate tests---see for instance \cite{Kay_1998, scharf94}. To our knowledge, however, the regime,  approach and results described in Section~\ref{sec2} are new.
	In the statistics community, the validation of testing procedures learning automatically the null distribution traces back to permutation and randomization tests. In the framework of multiple testing and of the FDR considered here, some results have been brought  recently, for instance in  \cite{BC2015,AC2017,RV2019}. 
	In the studies considered in Section \ref{sec3}, the validation of the approach is merely done by numerical simulations so far. One direction  of research is to obtain theoretical guarantees, possibly by modifying or simplifying the procedures and models, inspired by such recent results. 


	\bibliographystyle{unsrt}
	\bibliography{biblio}

\begin{thebibliography}{10}

\bibitem{BH1995}
Yoav Benjamini and Yosef Hochberg.
\newblock Controlling the false discovery rate: a practical and powerful
  approach to multiple testing.
\newblock {\em J. Roy. Statist. Soc. Ser. B}, 57(1):289--300, 1995.

\bibitem{Donoho_2004}
D.~{Donoho} and J.~{Jin}.
\newblock {Higher criticism for detecting sparse heterogeneous mixtures}.
\newblock {\em Ann. Stat.}, 2004.

\bibitem{Moscovich_2016}
A.~{Moscovich} et~al.
\newblock {On the exact Berk-Jones statistics and their p-value calculation}.
\newblock {\em Electron. J. Stat.}, 10:2329--2354, 2016.

\bibitem{Aldor_2013}
S.~{Aldor-Noiman} et~al.
\newblock The power to see: A new graphical test of normality.
\newblock {\em Am. Stat.}, 68(4):318--318, 2013.

\bibitem{Mary_2014}
D.~Mary and A.~Ferrari.
\newblock A non-asymptotic standardization of binomial counts in higher
  criticism.
\newblock In {\em Inform. Theory (ISIT), IEEE Int. Symp.}, pages 561--565, June
  2014.

\bibitem{Kaplan_2014}
D.M. {Kaplan} and M.~{Goldman}.
\newblock True equality (of pointwise sensitivity) at last: a dirichlet
  alternative to {K}olmogorov-{S}mirnov inference on distributions.
\newblock {\em Tech. report}, 2014.

\bibitem{Gontscharuk_2014}
V.~{Gontscharuk} et~al.
\newblock The intermediates take it all: Asymptotics of higher criticism
  statistics and a powerful alternative based on equal local levels.
\newblock {\em Biom. J.}, 57(1):159--180, 2014.

\bibitem{Dumusque_2012}
X.~{Dumusque} et~al.
\newblock {An Earth-mass planet orbiting {$\alpha$} Centauri B}.
\newblock {\em Nature}, 491:207--211, 2012.

\bibitem{Hatzes_2013}
A.~{Hatzes}.
\newblock {The Radial Velocity Detection of Earth-mass Planets in the Presence
  of Activity Noise: The Case of {$\alpha$} Centauri Bb}.
\newblock {\em ApJ}, 770:133, 2013.

\bibitem{Rajpaul_2016}
V.~{Rajpaul} et~al.
\newblock {Ghost in the time series: no planet for Alpha Cen B}.
\newblock {\em MNRAS}, 456:L6--L10, 2016.

\bibitem{sulis:tel-01687077}
Sophia Sulis.
\newblock {\em {Statistical methods using hydrodynamic simulations of stellar
  atmospheres for detecting exoplanets in radial velocity data}}.
\newblock Theses, {Universit{\'e} C{\^o}te d'Azur}, October 2017.

\bibitem{bacon2017}
R.~et~al. {Bacon}.
\newblock The muse hubble ultra deep field survey - i. survey description, data
  reduction, and source detection.
\newblock {\em A\&A}, 608:A1, 2017.

\bibitem{Mary_2020}
D.~Mary, R.~Bacon, S.~Conseil, L.~Piqueras, and A.~Schutz.
\newblock Origin: Blind detection of faint emission line galaxies in muse
  datacubes.
\newblock {\em Astronomy \& Astrophysics}, Jan 2020.

\bibitem{Perryman_2018}
M.~{Perryman}.
\newblock {\em The exoplanet handbook (2n edition)}.
\newblock Cambridge Univ., 2018.

\bibitem{Sulis20203DMS}
Sophia Sulis, D.~Mary, and L.~Bigot.
\newblock 3d magneto-hydrodynamical simulations of stellar convective noise for
  improved exoplanet detection. {I}. {C}ase of regularly sampled radial
  velocity observations.
\newblock {\em Astronomy \& Astrophysics, to appear}, 2020.

\bibitem{Gupta_2001}
S.K. {Gupta} et~al.
\newblock {UPSO three channel fast photometer}.
\newblock {\em Bulletin Astron. Soc. of India}, 29:479--486, 2001.

\bibitem{Trees_2002}
H.L. {Van Trees}.
\newblock {\em Optimum Array Processing: Part IV of Detection, Estimation, and
  Modulation Theory}.
\newblock John Wiley \& Sons, Inc., 2002.

\bibitem{ieee_sophia}
S.~{Sulis}, D.~{Mary}, and L.~{Bigot}.
\newblock A study of periodograms standardized using training datasets and
  application to exoplanet detection.
\newblock {\em IEEE Transactions on Signal Processing}, 65(8):2136--2150, April
  2017.

\bibitem{Li_2014}
T.H. {Li}.
\newblock {\em Time series with mixed spectra}.
\newblock CRC Press, 2014.

\bibitem{Fisher_1929}
R.A. {Fisher}.
\newblock {Tests of Significance in Harmonic Analysis}.
\newblock {\em Proc. R. Soc. London, Ser. A}, 125:54--59, 1929.

\bibitem{Chiu_1989}
S.T. Chiu.
\newblock Detecting periodic components in a white gaussian time series.
\newblock {\em J. R. Stat. Soc. Series B}, 51(2):249--259, 1989.

\bibitem{Shimshoni_1971}
M.~{Shimshoni}.
\newblock On fisher's test of significance in harmonic analysis.
\newblock {\em Geophys. J. R. Astronom. Soc.}, pages 373--377, 1971.

\bibitem{Bolviken_1983a}
E.~B{\"o}lviken.
\newblock New tests of significance in periodogram analysis.
\newblock {\em Scandinavian J. Stat.}, 10(1):1--9, 1983.

\bibitem{cheng2017}
Dan Cheng and Armin Schwartzman.
\newblock Multiple testing of local maxima for detection of peaks in random
  fields.
\newblock {\em Ann. Statist.}, 45(2):529--556, 04 2017.

\bibitem{Kay_1998}
S.M. {Kay}.
\newblock {\em Fundamentals of Statistical signal processing. Vol II :
  Detection theory.}
\newblock Prentice-Hall, Inc, 1998.

\bibitem{scharf94}
L.~L. Scharf and B.~Friedlander.
\newblock Matched subspace detectors.
\newblock {\em IEEE Trans. on Signal Processing}, 42(8):2146--2157, 1994.

\bibitem{BC2015}
Rina~Foygel Barber and Emmanuel~J. Cand\`es.
\newblock Controlling the false discovery rate via knockoffs.
\newblock {\em Ann. Statist.}, 43(5):2055--2085, 2015.

\bibitem{AC2017}
Ery Arias-Castro and Shiyun Chen.
\newblock Distribution-free multiple testing.
\newblock {\em Electron. J. Stat.}, 11(1):1983--2001, 2017.

\bibitem{RV2019}
Etienne {Roquain} and Nicolas {Verzelen}.
\newblock {On using empirical null distributions in Benjamini-Hochberg
  procedure}.
\newblock {\em arXiv e-prints}, page arXiv:1912.03109, Dec 2019.

\end{thebibliography}
	
	

\end{document}